\documentclass[aps,prc,nofootinbib,showpacs]{revtex4}

\usepackage{epsfig}
\usepackage{graphicx}

\begin{document}

\title{$\Theta^+$-pentaquark parity from associative baryon production in
$\pi D$ collisions, $\pi^{\pm}+D\to \Theta^+ +\Sigma^{\pm}$ }

\author{G. I. Gakh \footnote{E-mail: gakh@kipt.kharkov.ua} and A. P. Rekalo}
\affiliation{{\it National Science Centre "Kharkov Institute of Physics and Technology",\\ 61108 Akademicheskaya 1, Kharkov,
Ukraine}}
\author{E. Tomasi-Gustafsson}
\affiliation{\it DAPNIA/SPhN, CEA/Saclay, 91191 Gif-sur-Yvette Cedex, 
France}

\date{\today}


\begin{abstract}

We derive, in model independent way, the spin structure of the matrix element
for the reaction of associative $\Theta^+$  pentaquark production,
$\pi^{\pm}+D\to \Theta^+ +\Sigma^{\pm}, $ in the threshold region and in
collinear kinematics. The expressions for the polarization observables in
this reaction are found assuming spin 1/2 and different parities for
$\Theta^+ .$ We have proved that such reaction can be used for a model
independent determination of the P-parity of $\Theta^+$ only by measuring
the $\Theta^+$ polarization. Other polarization observables, such as the
dependence of the $\Sigma^{\pm}$ polarization on the vector and tensor
deuteron polarizations, are insensitive to the $\Theta^+$ parity in the 
considered kinematical conditions.  All linear and quadratic
relations between polarization observables in $\pi^{\pm}+\vec D\to
\Theta^+ +\vec\Sigma^{\pm}(\Theta^+ $ is unpolarized) do not depend on the
parity of the $\Theta^+$ pentaquark.
\end{abstract}
\vspace{0.2cm}

\maketitle
\section{Introduction}
The possible existence of exotic hadrons was suggested in connection with
the $KN$ scattering data \cite{G} (before the advent of QCD). In QCD the
multiquark states were considered as natural extension of the ordinary
hadrons. A state with mass about 1.5 GeV and a strong decay width
smaller than 15 MeV, now called $\Theta^+$ pentaquark,  was predicted in Ref. \cite{DPP}. This work motivated and oriented 
experimental searches recently leading  to the observation of a narrow
resonance at about the predicted mass \cite{NB}. So far this observation
has been confirmed or infirmed by other experimental groups using various projectiles
and targets (for the details see the review \cite{S}).

The narrow exotic (B=1, S=1) baryon resonance $\Theta^+ (1540)$ was called
pentaquark since the simplest quark content of $\Theta^+$ is ($uudd\bar s$).
The quantum numbers of the pentaquarks are very important for the determination of its quark structure and in particular its multiplet assignment \cite{J}.

At present the existence of a narrow pentaquarks is not fully confirmed:
several laboratories have reported the evidence of such states, while others
have negative results. The question of whether pentaquarks exist may be solved by a second generation of high statistics experiments (for example,
at JLab \cite{K}).

The parity of the $\Theta^+$ pentaquark, $\pi(\Theta)$, being a very
important characteristics of this exotic state, is especially challenging
for a model independent determination. The study of polarization
phenomena for different processes of $\Theta^+$ production seems unavoidable.
Unfortunately, in many cases, the necessary set of observables includes the
$\Theta^+$ polarization which measurement, through the
decay $\Theta^+\to NK$, is not an easy task.

Up to now, the most perspective method for the determination of
$\pi(\Theta)$, in model independent way, has been suggested for associative
hyperon production in nucleon-nucleon collisions, $N+N\to Y+\Theta^+$,
$Y=\Lambda$ or $\Sigma$, and $p+p\to \Theta^+ +\Lambda +\pi ^+ $ (\cite{Re04} and
refs therein). It was shown, that in the threshold region, (with s-wave
production of the final hyperons) double spin polarization observables, such
as the spin correlation coefficients $C_{xx}$ or $C_{yy}$ (in the collision of
polarized nucleons), and the spin transfer coefficient $D_{xx}$ (from one
initial nucleon, beam or target, to the produced hyperon $Y$) is sensitive
to $\pi(\Theta)$.

We discuss in the present work another possible, simple reaction, the two
body associative hyperon production, with baryon number equal 2, similar to
$N+N\to  \Theta^+ +Y$, but with a deuteron in the initial state:
\begin{equation}\pi^{\pm}+D\to \Theta^+ + \Sigma^{\pm} \label{eq:eq1}
\end{equation}
in the threshold region [$E_{\pi ,th}$= 1.04 GeV, with $M(\Theta^+)$=1.54
GeV] or in collinear kinematics, with the aim of finding some polarization
observables which are sensitive to $\pi(\Theta)$. Note that a similar
reaction, $\pi^-+D\to n+n$, has been suggested many years ago \cite{Ch54}
for the model independent determination of the charged pion parity.

The study of the reaction (\ref{eq:eq1}) in collinear kinematics  looks
more preferable in comparison with the threshold region, due to the larger
cross section in forward direction, and to the fact that the formalism
applies to any pion energy.

\section{The reaction $\pi^{\pm}+D\to \Theta^+ + \Sigma^{\pm}$ }

\begin{figure}
\mbox{\epsfxsize=10.cm\leavevmode \epsffile{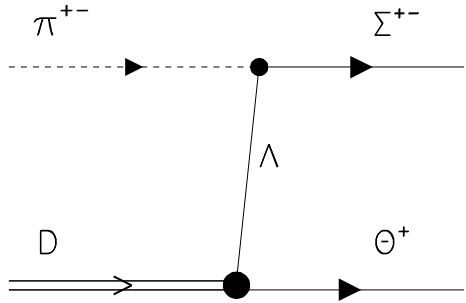}}
\caption{ Possible mechanism for $\pi^{\pm}+D\to \Theta^+ + \Sigma^{\pm} $.}
\label{fig:fig1}
\end{figure}

The simplest  mechanism for the reaction (\ref{eq:eq1}) is shown in Fig.
\ref{fig:fig1}. It is similar to the impulse approximation, where the new
deuteron vertex, $ D\to\Theta^+ + \Lambda$, generates a deuteron component
made from eight quarks: $D\to 3u+3d+s\bar s$. This component is not new.
For example, the presence of an intrinsic $s\bar s$ component in
the nucleon has been advocated to explain the strong violation of the OZI rule,
which has been observed in different processes of $\phi$ production, in
particular in $p\bar p$ annihilation at rest ( for a recent review see
Ref. \cite{No02}). It is possible to rearrange the $np$ system, composed by
$(uud+s\bar s)+(udd)$, as superposition of $\Theta^+$ and $  \Lambda$
hyperons and to derive a connection between the deuteron structure at short
distances and the $\Theta^+$ physics.
However, our interest here is focused on the dependence of the polarization
effects in the reaction (\ref{eq:eq1}) on the parity of the $\Theta^+ $
hyperon, $\pi(\Theta)$. To solve this problem we use only general symmetry
properties of the strong interaction, such as the P-invariance and the
isotropy of space.

In the general case, the matrix element for $\pi^{\pm}+D\to \Theta^+
+ \Sigma^{\pm} $ contains six independent spin structures, assuming spin 1/2
for $\Theta^+$, which have different parametrizations according to
$\pi(\Theta)$. From the isotopic invariance we have $\pi(p)=\pi(n)$.
The capture process, $\pi^-+D\to n+n$ \cite{Ch54}, proved the pseudoscalar
nature of the pion. In this case no polarization observable is necessary, as
the identity of the produced neutrons generates only one spin structure.
Unfortunately, one can not use such procedure for $\pi^{\pm}+D\to \Theta^+
+ \Sigma^{\pm} $, due to different final hyperons.

To avoid the complexity of the calculation of the polarization phenomena in
reaction (\ref{eq:eq1}), we restrict our analysis to two different kinematical
regimes: threshold production and collinear regime. Although the dynamics is
different in these two cases, the analysis of polarization effects is
similar, due to the axial symmetry of the two regimes: only one
physical direction is defined: the three-momentum of the colliding particles.
This essentially simplifies the calculations. Let us start from
considerations which hold in threshold regime.

\section{Matrix element and polarization effects}

The production of the final hyperons in $s$-state is characterized by two
different partial transitions:
\begin{eqnarray}
\underline{\pi(\Theta)=+1:} ~~~\ell=1&\to& {\cal J}^P=0^+\to S_f=0,\nonumber
\\ \ell=1&\to&{\cal J}^P=1^+\to S_f=1,\label{eq:tr1}
\end{eqnarray}
\begin{eqnarray}
\underline{\pi(\Theta)=-1:}~~~ \ell=0&\to& {\cal J}^P=1^-\to S_f=1,\nonumber
\\ \ell=2&\to&{\cal J}^P=1^-\to S_f=1,\label{eq:tr2}
\end{eqnarray}
where $\ell$ is the orbital angular momentum of the initial pion (in the
reaction CMS), $S_f$ is the total spin of the the $\Theta\Sigma$-system,
$ {\cal J}^P$ is the total angular momentum and parity of colliding (and
final) particles.

The spin structure of the corresponding matrix element can be written as:
\begin{eqnarray}
{\cal M}^{(+)}=\chi_2^{\dagger} \left [f_1^{(+)}\vec\sigma\cdot\vec U\times
\hat{\vec q} + i f_2^{(+)}\hat{\vec q}\cdot\vec U\right
]\sigma_y\tilde\chi_1^{\dagger},\label{eq:m1}\\
{\cal M}^{(-)}=\chi_2^{\dagger} \left [f_1^{(-)}\vec\sigma\cdot\vec U+
f_2^{(-)}\vec \sigma\cdot\hat{\vec q}\hat{\vec q}\cdot\vec U\right
]\sigma_y\tilde\chi_1^{\dagger},\label{eq:m2}
\end{eqnarray}
where $\chi_1$ and $\chi_2$ are the two component spinors of the $\Theta^+$
and $\Sigma$ hyperon, $\hat{\vec q}$ is the unit vector along the pion three
momentum, $\vec U$ is the deuteron polarization three-vector,
$f_i^{(\pm)}=f_i^{(\pm)}(w)$, $i=1,2$ are the partial amplitudes of the
transitions (\ref{eq:tr1}) and (\ref{eq:tr2}), which are, in general case,
complex functions of the excitation energy $w=\sqrt{s}-M_1-M_2$, $M_{1,2}$
are the masses of $\Theta$ and $\Sigma$, $\sqrt{s}$ is the invariant total
energy of the colliding particles.

Note that the spin structure of the matrix elements ${\cal M}^{(\pm)}$ in
collinear regime (i.e., for any value of $\sqrt{s}$ and production angle
equal $0^0$ or $180^0$) is described by the same formulas, Eqs. (\ref{eq:m1}) and
(\ref{eq:m2}), but the amplitudes $f_i^{(\pm)}\to f^{(\pm)}_{i,col}$ describe
two different allowed helicity transitions, which work in collinear
kinematics. Therefore, so different kinematical conditions are described by
different amplitudes, but, due to the axial symmetry inherent to the
considered problem, the same analysis of polarization phenomena can be
applied.

The axial symmetry and the P-invariance of the strong interaction allow
only single spin polarization observables in $\pi+\vec D\to  \Theta^+ +
\Sigma$, namely, the analyzing power induced by the quadrupole polarization
of the deuteron target:
\begin{equation}
\displaystyle\frac{d\sigma}{d\Omega}=\left (\displaystyle\frac{d\sigma}
{d\Omega}\right)_0\left [ 1+{\cal A} Q_{ab}\hat q_a\hat q_b\right ],
\label{eq:eqa}
\end{equation}
where ${\cal A}$ is the analyzing power, and $Q_{ab}$ is the tensor
which describes the deuteron quadrupole polarization. The spin--density
matrix of the deuteron is described by (in the reaction CMS):
\begin{equation}
\rho_{ab}=\displaystyle\frac{1}{3}\left ( \delta_{ab}
+\frac{q_aq_b}{M^2}\right )-\displaystyle\frac{i}{2}
\epsilon_{abc}\tilde S_c+Q_{ab},
\label{eq:rho}
\end{equation}
$$ Q_{aa}=\frac{q_aq_b}{E^2}Q_{ab},~ Q_{ab}= Q_{ba}, \ \
\tilde S_i=\frac{E}{M}S_i-\frac{{\vec q}\cdot {\vec S}}{ME}q_i,$$
where $\vec S$ and $E(M)$ are the vector polarization and energy (mass) of
the deuteron target in the reaction CMS.

The dependence of the $\Sigma$-hyperon polarization ${\vec P}_{\Sigma }$, on
the polarization characteristics of the deuteron target, can be written in
the following general form:
\begin{equation}
{\vec P}_{\Sigma }=\vec S {\cal P}_1+\hat{\vec q} \vec S \cdot \hat{\vec q}
{\cal P}_2+\vec Q\times \hat{\vec q}{\cal P}_3,
\label{eq:pol}
\end{equation}
$$ Q_{a}= Q_{ab}\hat{ q}_b,$$
where ${\cal P}_i$, $i=1-3$, are real functions which can be expressed in
terms of bilinear combinations of the reaction amplitudes. These functions
are related to the following coefficients of the polarization transfer from
the initial deuteron to the produced $\Sigma$-hyperon:
$$D_{xx}=D_{yy}={\cal P}_1,~D_{zz}={\cal P}_1+{\cal P}_2,
~D_{x,yz}=-D_{y,xz}={\cal P}_3,$$
where the $z$-axis is chosen along the $\hat{\vec q}$ direction, but the
$x$ and $y$ directions can not be uniquely fixed, due to the axial symmetry.

Summarizing, it is possible to measure four T-even observables:
$ ({d\sigma}/{d\Omega})_0,$~${\cal P}_1$, ${\cal P}_2$ and ${\cal A} $ and
one T-odd observable, the coefficient ${\cal P}_3$.

Therefore, we can state that there must be a definite linear relations
between T-even observables, which is model independent and works in
collinear and threshold regime, as well.

To find this relation, let us express the coefficients ${\cal A} $,
${\cal P}_1$, ${\cal P}_2$ and  ${\cal P}_3$ as quadratic combinations of
the corresponding amplitudes (for both cases of the $\Theta^+$ parity):

$\noindent \underline{\pi(\Theta)=+1:} $
$$\left (\displaystyle\frac{d\sigma_+}{d\Omega}\right)_0=
\displaystyle\frac{N}{3}\left (2|f_1^{(+)}|^2+r^2 |f_2^{(+)}|^2 \right),
~{\cal A}^{(+)}\left (\displaystyle\frac{d\sigma_+}{d\Omega}\right)_0=
\displaystyle N\left (-r^{-2}|f_1^{(+)}|^2+ |f_2^{(+)}|^2\right ), $$
$${\cal P}_1^{(+)}\left (\displaystyle\frac{d\sigma_+}{d\Omega}\right)_0=
-NrRe f_1^{(+)}f_2^{(+)*},~{\cal P}_2^{(+)}\left (\displaystyle\frac{d\sigma_+}
{d\Omega}\right)_0=Nr\left (r^{-2}|f_1^{(+)}|^2+Re f_1^{(+)}f_2^{(+)*}
\right ),~ $$
$${\cal P}_3^{(+)}\left (\displaystyle\frac{d\sigma_+}{d\Omega}\right)_0=
2NIm f_1^{(+)}f_2^{(+)*}, $$
$\underline{\pi(\Theta)=-1:} $
$$\left (\displaystyle\frac{d\sigma_-}{d\Omega}\right)_0=\displaystyle
\frac{N}{3}\left (2|f_1^{(-)}|^2+ r^2|f_1^{(-)}+f_2^{(-)}|^2\right ),
~{\cal A}^{(-)}\left (\displaystyle\frac{d\sigma_-}{d\Omega}\right)_0=
\displaystyle N\left (-r^{-2}|f_1^{(-)}|^2+ |f_1^{(-)}+f_2^{(-)}|^2\right ), $$
$${\cal P}_1^{(-)}\left (\displaystyle\frac{d\sigma_-}{d\Omega}\right)_0=
Nr\left (|f_1^{(-)}|^2+Ref_1^{(-)}f_2^{(-)*}\right ),
~{\cal P}_2^{(-)}\left (\displaystyle\frac{d\sigma}{d\Omega_-}\right)_0=
-Nr\left (Re f_1^{(-)}f_2^{(-)*}+\frac{{\vec q}^2}{E^2}|f_1^{(-)}|^2\right ),$$
$${\cal P}_3^{(-)}\left (\displaystyle\frac{d\sigma_-}{d\Omega}\right)_0=-
2NIm f_1^{(-)}f_2^{(-)*}, $$
where $r=E/M$ and $N= (1/64\pi^2)(p/q_{\pi})(1/MW)$ is a kinematical factor.

From these formulas, one can find a linear relation between T-even polarization
observables:
\begin{equation}
{\cal P}_1^{(\pm)}+{\cal P}_2^{(\pm)}+\frac{r}{3}{\cal A}^{(\pm)}=
r^{-1}~\mbox{ or~}
D_{zz}^{(\pm)} +\frac{r}{3}{\cal A}^{(\pm)}=r^{-1}.
\label{eq:eq8}
\end{equation}
This relation is the same for the two possible values of the $\Theta^+$ parity.
Therefore, it can not be used for the determination of this parity. Being 
model independent, this relation can be useful to determine the extension of
the threshold region, i.e., of the $s$-wave dominance.

The measurement of the analyzing power ${\cal A}$ and cross section
$ ({d\sigma}/{d\Omega})_0$ allows to realize the first step of the complete
experiment for $\pi^{\pm}+D\to \Theta^+ +\Sigma^{\pm} $, i.e., the
determination of the moduli of the two complex amplitudes:
$$N|f_1^{(+)}|^2=(1-\frac{r^2}{3}{\cal A}^{(+)})\left
(\displaystyle\frac{d\sigma_+}{d\Omega}\right)_0,
Nr^2~|f_2^{(+)}|^2=(1+\frac{2r^2}{3}{\cal A}^{(+)})\left
(\displaystyle\frac{d\sigma}{d\Omega_+}\right)_0, $$
\begin{equation}
N|f_1^{(-)}|^2=(1-\frac{r^2}{3}{\cal A}^{(-)})\left (\displaystyle
\frac{d\sigma_-}{d\Omega}\right)_0,
~Nr^2|f_1^{(-)}+f_2^{(-)}|^2=(1+\frac{2r^2}{3}{\cal A}^{(-)})\left (
\displaystyle\frac{d\sigma_-}{d\Omega}\right)_0.
\label{eq:eq9}
\end{equation}
Evidently, the T-odd observable ${\cal P}_3$ , being proportional to
$\sin\delta$ ($\delta$ is the relative phase of the complex amplitudes
$f_1^{(\pm)}$ and $f_2^{(\pm)}$), is the most sensitive to this phase.
This is the last step of the complete experiment. It is possible also to
find definite quadratic relations between T-odd and T-even observables:
\begin{equation}
1+\frac{r^2}{3}{\cal A}^{(\pm)}-\frac{2r^4}{9} ({\cal A}^{(\pm)})^2-
\displaystyle\frac{r^2}{4}({\cal P}_3^{(\pm)})^2=({\cal P}_1^{(\pm)})^2,
~\mbox{ if~} \pi(\Theta)=\pm 1.
\label{eq:eq10}
\end{equation}
It is another result, which shows also that the specific quadratic
combination of different observables of T-odd and T-even nature does not
depend on $\pi(\Theta)$. In order to test these relations, it necessary to
have a deuteron target, vector and tensor polarized, and to measure the
$\Sigma$-hyperon polarization. These linear and quadratic relations apply
equally well for threshold and for collinear kinematics of the considered
process.

But the situation is different for the  $\Theta^+$ polarization. The
dependence of the $\Theta^+$ pentaquark polarization ${\vec P}_{\Theta }$,
on the polarization characteristics of the deuteron target, can be written
in the following general form (similar to Eq. (8)):
\begin{equation}
{\vec P}_{\Theta }=\vec S \bar{\cal P}_1+\hat{\vec q} \vec S \cdot \hat
{\vec q}\bar{\cal P}_2+\vec Q\times \hat{\vec q}\bar{\cal P}_3,
\label{eq:pol1}
\end{equation}
where $\bar{\cal P}_i$, $i=1-3$, are real functions which can be expressed
in terms of bilinear combinations of the reaction amplitudes. Using Eqs. (4)
and (5) for the matrix elements we can obtain the following expressions for
$\bar P_i$ in terms of quadratic combinations of the corresponding amplitudes
(for both cases of the  $\Theta^+$ parity)

$\noindent \underline{\pi(\Theta)=+1:} $
$${\cal P}_1^{(+)}=-\bar {\cal P}_1^{(+)}, \ \
\bar {\cal P}_2^{(+)}\left (\displaystyle\frac{d\sigma_+}
{d\Omega}\right)_0=Nr\left (r^{-2}|f_1^{(+)}|^2-Re f_1^{(+)}f_2^{(+)*}
\right ), \ \  {\cal P}_3^{(+)}=-\bar {\cal P}_3^{(+)}, $$
$\noindent \underline{\pi(\Theta)=-1:} $
$${\cal P}_i^{(-)}=\bar {\cal P}_i^{(-)}, \ i=1, 2, 3. $$
From these formulas one can see that only the transversal components of the
$\Theta^+$ polarization vector are sensitive to the pentaquark parity:
$$\bar {\cal P}_{1,3}=-\pi(\Theta^+){\cal P}_{1,3},~\bar{\cal P}_1+\bar
{\cal P}_2={\cal P}_1+{\cal P}_2.$$
In terms of the coefficients of polarization transfer $D_{ab}(\Theta)$
[$D_{ab}(\Sigma)$] from vector--polarized deuteron target to the $ \Theta^+$
($\Sigma$) hyperons, these relations can be written as follows:
$$D_{aa}(\Theta)=-\pi(\Theta)D_{aa}(\Sigma),~aa=xx; yy; x,yz ~\mbox{ or~}y,xz,
~D_{zz}(\Theta)=D_{zz}(\Sigma).$$
Evidently, it is a very difficult experiment.

\section{Conclusions}

We performed a model independent analysis of the spin structure of the
matrix element for associative hyperon production in $\pi^{\pm}D$ collisions,
$\pi^{\pm}+D\to \Theta^+ +\Sigma^{\pm} $, in collinear or threshold regime.
We calculated different polarization observables and showed that they are
related by very specific linear and quadratic equations. Unfortunately, these
model independent relations are insensitive to the parity of the
$\Theta$-pentaquark. Therefore, difficult measurements of the $\Theta^+$
polarization are necessary in order to determine the parity of the
$\Theta^+$.

What is the physical reason of such insensitivity of the $\pi^{\pm}+D\to
\Theta^+ +\Sigma^{\pm} $ process to $\pi(\Theta)$? What is the main
difference between $N+N\to Y+\Theta$ and  $\pi^{\pm}+D\to \Theta^+
+\Sigma^{\pm} $, which look very similar processes? The problem is that the
initial two-nucleon system in $N+N\to Y+\Theta$  is more flexible, as it
can be in triplet and singlet state, as well, whereas in  $\pi^{\pm}+D\to
\Theta^+ +\Sigma^{\pm} $ the initial two-nucleon system, being a deuteron,
is always in triplet state.

However, if the parity of $\Theta^+$ will be determined, the study of the
process $\pi^{\pm}+D\to \Theta^+ +\Sigma^{\pm} $ will be useful for the
study of other questions, related to the $\Theta^+$ physics, as, for example,
the exotic baryon content of the deuteron at small distances.

{}


\begin{thebibliography}{}
\bibitem{G}
E. Golowich, Phys. Rev. {\bf D4}, 262 (1971).
\bibitem{DPP}
D. Diakonov, V. Petrov M. Polyakov, Z. Phys. {\bf A359}, 305 (1987).
\bibitem{NB}
T. Nakano et al., Phys. Rev. Lett. {\bf 91}, 012002 (2003);
V. V. Barmin et al., Phys. Atom. Nucl. {\bf 66}, 1715 (2003)
[Yad. Phys. {\bf 66}, 1763 (2003)].
\bibitem{S}
Fl. Stancu, hep-ph/0408042.
\bibitem{J}
  R.~Jaffe and F.~Wilczek,
  Eur.\ Phys.\ J.\ C {\bf 33}, S38 (2004).
\bibitem{K}
  V.~Kubarovsky and P.~Stoler  [CLAS Collaboration],
  Nucl.\ Phys.\ Proc.\ Suppl.\  {\bf 142}, 356 (2005).
\bibitem{Re04}
M.~P.~Rekalo and E.~Tomasi-Gustafsson,
  Eur.\ Phys.\ J.\ A {\bf 22}, 119 (2004) and Refs herein.
\bibitem{Ch54} W. Chinowsky and J. Steinberger, Phys. Rev. 95, 1561 (1954).

\bibitem{No02}
  V.~P.~Nomokonov and M.~G.~Sapozhnikov,
  Phys.\ Part.\ Nucl.\  {\bf 34}, 94 (2003)
  [Fiz.\ Elem.\ Chast.\ Atom.\ Yadra {\bf 34}, 189 (2003)].


\end{thebibliography}
\end{document}